\documentclass[10pt,twocolumn]{IEEEtran}
\pdfoutput=1

\usepackage{amsmath,amssymb}
\usepackage{graphicx}
\usepackage{booktabs}
\usepackage{tikz}
\usetikzlibrary{positioning,arrows.meta,calc}
\usepackage[hidelinks]{hyperref}

\title{Closed-Loop Bayesian Bandit Encoder\\
with GRAND Receiver for a Bursty Interference Channel}

\author{Bhaskar Krishnamachari\\
Ming Hsieh Department of Electrical and Computer Engineering\\
Viterbi School of Engineering, University of Southern California\\
bkrishna@usc.edu}

\begin{document}
\maketitle

\begin{abstract}
Interleaving mitigates burst errors but introduces decoding delay and
removes temporal error structure that a channel-aware decoder could exploit.
We consider packet-level selection between a random linear code and the same
code used with cross-codeword interleaving, over a channel with an unknown
number of on/off interferers. The receiver uses Guessing Random Additive
Noise Decoding (GRAND) with a replaceable
noise model and feeds aggregate channel statistics back to a Bayesian
estimator at the transmitter. Once the interference amplitudes and timing
parameters are estimated, the receiver's noise model is replaced: it
computes hidden-Markov-model posterior bit-flip probabilities and uses them
to order GRAND queries. A discounted Thompson sampler selects
between the two transmission modes using a goodput-minus-latency reward
whose distribution is endogenously nonstationary: receiver adaptation,
rather than channel change, alters the value of each mode.
Across five simulation seeds, the interleaved mode is preferred before
channel estimation converges. After the learned decoder is activated, the
non-interleaved mode becomes preferable because it achieves lower block
error rate without interleaving delay. In the reference configuration, the
learned noise model reduces block error rate by approximately one order of
magnitude relative to ORBGRAND. Using partial channel estimates before full
convergence reduces pre-convergence block error rate by up to $4.5\times$.
Adding model-predicted utilities as confidence-weighted pseudo-observations
reduces post-transition selection of the inferior arm by approximately
$65\%$. Under an idealized airtime conversion at a 100~MHz 5G~NR-like
symbol rate, the learning transient corresponds to a few milliseconds of
occupied symbol time.
\end{abstract}

\section{Introduction}
\label{sec:intro}

Interleaving is commonly used to convert burst errors into approximately
independent errors before decoding. This improves the performance of
decoders designed for memoryless channels, but it introduces block delay and
discards temporal error structure. If the decoder can instead learn and
exploit that structure, the benefit of interleaving may disappear while its
delay cost remains.

Guessing Random Additive Noise Decoding (GRAND) \cite{duffy2019grand} is
useful in this setting because its codebook test and its noise model are
modular. The code determines only the
membership query, while the channel model determines the order in which
candidate noise patterns are tested. A receiver can therefore update the
noise model online without changing the code, the encoder, or the hardware
that checks codebook membership.

This report studies a transmitter that chooses, for each packet, between a
non-interleaved random linear code and the same code used with a
cross-codeword interleaver. The receiver initially uses a generic noise
model; a learner at the transmitter estimates the interference process from
aggregate receiver feedback, and the receiver later switches to a
burst-aware model built from that estimate. Because this update changes
the relative performance of the two transmission modes, code selection is
formulated as a two-armed bandit whose nonstationarity is
\emph{endogenous}: the reward distribution of the non-interleaved arm
changes not because the channel changes but because the receiver's own
learning changes the decoder. This coupling between learning at one end of
the link and decision-making at the other is the defining feature of the
problem. We hypothesize that the
preferred arm changes after receiver adaptation: the interleaved arm is
preferable initially, whereas the non-interleaved arm is preferable after
the learned noise model is activated, since it then achieves lower error
rate without the interleaving delay.

The main contributions are:
\begin{enumerate}
\item A closed-loop architecture that combines online interference
estimation, runtime adaptation of a GRAND noise model, and packet-level
selection between interleaved and non-interleaved transmission, coupled so
that decoder adaptation itself induces the nonstationarity the
code-selector must track.
\item A channel-estimation and decoding method that estimates interferer
amplitudes and timing from thresholded run statistics with an analytic
correction for threshold-induced bias, computes HMM-smoothed bit-flip
probabilities, and uses those probabilities to order GRAND queries.
\item A simulation study that characterizes when the preferred transmission
mode changes, and that compares receiver gating policies and ACK-only
versus model-informed Thompson sampling.
\end{enumerate}

To the best of our knowledge, prior work has not treated interleaved and
non-interleaved realizations of one code as bandit arms, nor studied the
case in which the reward distribution of one arm changes mid-connection
because the receiver's decoder is being retuned by a channel learner.

The remainder of this report is organized as follows.
Section~\ref{sec:grand} reviews GRAND. Section~\ref{sec:related} discusses
related work. Section~\ref{sec:model} presents the system and channel
model. Section~\ref{sec:estimation} describes online interference
estimation, and Section~\ref{sec:noisemodel} the learned noise model and
gating policies. Section~\ref{sec:bandit} describes the bandit.
Section~\ref{sec:experiments} details the experimental phases.
Section~\ref{sec:results} presents results, and
Section~\ref{sec:timescales} gives an idealized airtime conversion for a
5G~NR-like link. Section~\ref{sec:conclusion} concludes with limitations
and future work.

\section{Background: GRAND}
\label{sec:grand}

Consider a linear block code of length $n$ and dimension $k$, and let $y^n$
be the demodulated channel output. Classical decoding searches the codebook
for the codeword closest to $y^n$ under the channel metric. GRAND
\cite{duffy2019grand} inverts the search: it guesses the \emph{noise}. The
receiver generates putative noise-effect sequences $z^n$ in decreasing order
of likelihood, subtracts each from the hard-detected word, and queries
whether the result is a codebook member; for a linear code the query is a
syndrome check. The first hit is provably a maximum likelihood (ML)
decoding, provided the query order matches the channel statistics. Three
properties matter for this work. First, GRAND is code-agnostic: the code
enters only through membership queries, so any code of moderate redundancy
can be decoded, and random linear codes (RLCs) with no algebraic structure
become practical. Second, the expected number of queries scales with the
redundancy $n-k$ rather than with the codebook size, so high-rate codes are
the practical operating regime. Third, the channel model is confined to the
pattern generator, so a receiver can swap noise models at runtime.

GRANDAB \cite{duffy2019grand} abandons decoding after a query budget $Q$
and declares an erasure; the budget bounds both complexity and, as verified
in Section~\ref{sec:results}, the achievable block error rate through the
tail probability that the true noise pattern lies beyond the queried set.
For soft detection, ORBGRAND \cite{duffy2022orbgrand} sorts bits by
reliability $|{\rm LLR}|$ and generates patterns in increasing
\emph{logistic weight}, the sum of the flipped bits' reliability ranks; it
is hardware-friendly and almost capacity-achieving \cite{liu2023orbgrand},
with fabricated decoders demonstrating multi-Gb/s operation
\cite{abbas2022vlsi,riaz2023isscc}. SGRAND \cite{solomon2020sgrand} instead
uses the full per-bit soft values and a max-heap to emit patterns in
decreasing probability under an independent per-bit error model. GRAND has
also been extended to channels with memory: GRAND-MO \cite{an2022bursts}
orders patterns for a Markov error process and demonstrates that abandoning
interleaving can simultaneously improve error rate and latency when errors
arrive in dense bursts. Our learned noise model
(Section~\ref{sec:noisemodel}) combines the SGRAND ordering machinery with
per-bit posteriors computed from a learned interference model, which
handles the sparser burst errors that arise at moderate interferer
amplitudes.

\section{Related Work}
\label{sec:related}

\textit{Noise-guessing decoding.}
GRAND reframes decoding as an ordered search over candidate noise sequences,
with the code entering only through a codebook-membership test
\cite{duffy2019grand}. We therefore use \emph{code-agnostic} to describe this
property, reserving \emph{channel-universal} for decoders that do not know
the channel law. Soft-information variants progressively refine the query
order: SRGRAND uses quantized symbol reliability \cite{duffy2022srgrand},
ORBGRAND uses reliability ranks and logistic weight
\cite{duffy2022orbgrand,liu2023orbgrand}, and SGRAND orders patterns by their
exact likelihood under independent per-bit error probabilities
\cite{solomon2020sgrand}. GRAND-MO exploits Markov error memory and showed
that retaining bursts can outperform interleaving when the burst model is
known \cite{an2022bursts}. Recent soft-detection work extends this
interleaving-free line to further channels with memory: GRAND has been
adapted to Gaussian intersymbol-interference channels \cite{li2026isi}, and
ORBGRAND-AI decodes in the presence of ISI without interleaving
\cite{duffy2025orbgrandai}. Hardware implementations of ORBGRAND
demonstrate that noise-guessing decoding can support multi-Gb/s operation
\cite{abbas2022vlsi,riaz2023isscc}. Most of these systems assume that the
query-order model is fixed or known. Recent channel-universal
noise-guessing decoders estimate unknown finite-state additive laws within
the decoding rule \cite{miyamoto2025universal}, while joint
channel-estimation/GRAND methods account for fading-estimation errors
\cite{wiame2025joint}. All of these works adapt the decoder to a memory
channel whose law is known, assumed, or estimated within the decoding rule;
none couples that adaptation to a transmitter-side decision. The
distinguishing element of the present work is therefore not decoding
without interleaving per se, which
\cite{an2022bursts,li2026isi,duffy2025orbgrandai} establish for fixed or
locally estimated channels, but the closed loop: interference parameters
are learned at the transmitter from aggregate receiver feedback, the
estimate replaces GRAND's noise model during the connection, and a bandit
co-adapts the transmission mode, so that the reward process facing the
transmitter is endogenously nonstationary.

\textit{Bursty channels and interleaving.}
The Gilbert and Elliott models are the classical two-state descriptions of
burst errors \cite{gilbert1960,elliott1963}, and finite-state channels with
memory have long been used to analyze coding and interleaving
\cite{kanal1978}. Interleaving makes the residual errors presented to a
memoryless decoder closer to independent, but it also introduces block delay
and removes temporal structure that a matched decoder could exploit. The
closest coding result to ours is GRAND-MO, which establishes this
reliability--latency advantage for a fixed, known Markov noise process
\cite{an2022bursts}. Our setting differs in two respects: the interference is
a superposition of multiple on/off sources with unknown amplitudes and
timing, and at moderate amplitudes its correlation is more visible in the
soft samples than in hard-decision error runs. This motivates learning
physical interference parameters and forming HMM-smoothed bit-flip posteriors
rather than fitting a fixed Gilbert--Elliott error process.

\textit{Channel learning and adaptive receivers.}
Cognitive radio established the broader principle of learning the radio
environment and adapting transmission decisions \cite{haykin2005}, while
finite-state and hidden-state channel models provide tools for inference when
the disturbance has memory \cite{kanal1978}. Existing adaptive receivers
typically use channel estimates to update equalization, demodulation, or
decoder reliabilities; recent GRAND work similarly couples fading-channel
uncertainty to the decoding metric \cite{wiame2025joint}. Our learner instead
infers the number, amplitudes, duty cycles, and dwell times of interfering
sources from compact aggregate feedback. The resulting model has two
simultaneous uses: it changes the receiver's ordering of candidate noise
patterns and predicts the utility of alternative transmitter-side code
structures. This shared model is what closes the loop between channel
learning, decoder adaptation, and code selection.

\textit{Bandits for link adaptation.}
Multi-armed bandits are well established for cognitive channel access
\cite{lai2011access} and for selecting transmission rates or
modulation-and-coding schemes \cite{combes2019ors,qi2019rate,saxena2020bayesla}.
Thompson sampling provides a Bayesian exploration mechanism
\cite{thompson1933,russo2018tutorial}, and discounting or sliding windows are
standard responses to changing reward distributions
\cite{garivier2011switching}. Very recent work further exploits structural
dependence among MCS arms through a joint Thompson-sampling prior
\cite{vinjam2026jointts}. These studies choose among channels, rates, or MCS
levels whose success probabilities are unknown or time-varying. Here the arms
are two realizations of the same code, non-interleaved and cross-codeword
interleaved, and the nonstationarity is partly endogenous: learning at the
receiver changes the decoder, which selectively changes the reward
distribution of the non-interleaved arm. The latency-aware reward and the use
of model-predicted utilities as confidence-weighted pseudo-observations are
designed for this closed-loop setting. Thus, the closest prior works address
noise-model adaptation, burst-aware decoding, and bandit link selection
separately; this work studies their interaction in a single online system.

\section{System and Channel Model}
\label{sec:model}

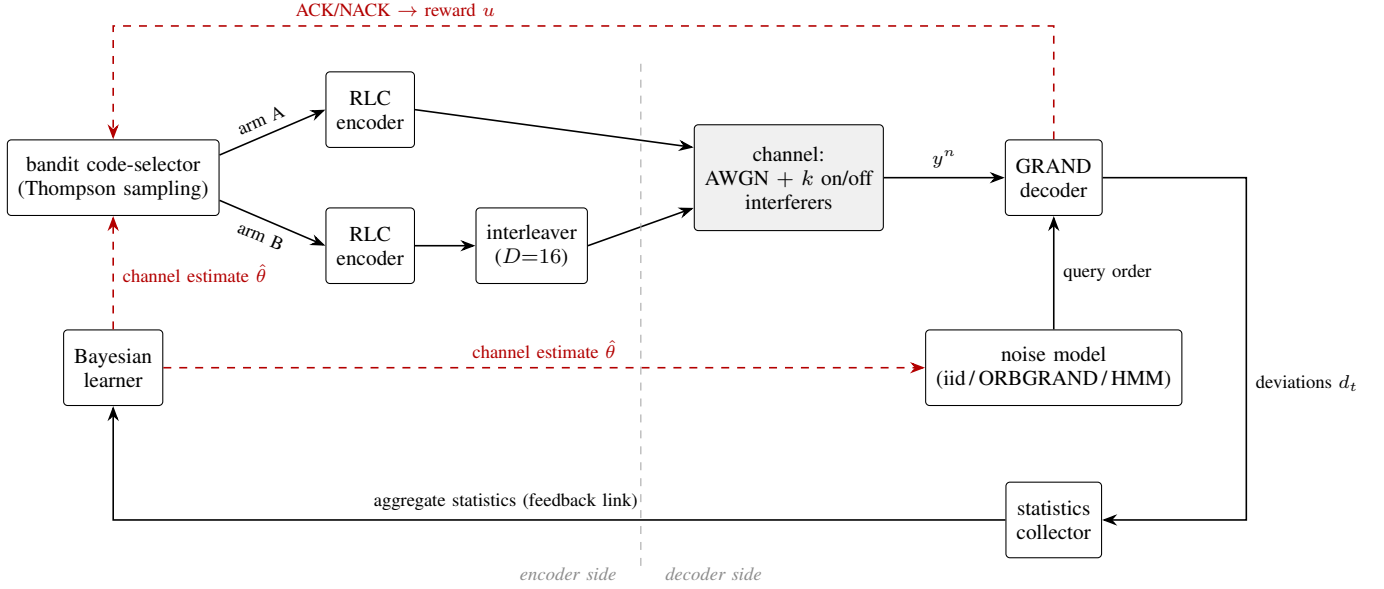
\begin{figure*}[t]
\centering
\begin{tikzpicture}[
  box/.style={draw, rounded corners=1.5pt, align=center, font=\footnotesize,
              inner sep=4pt, minimum height=10mm},
  lbl/.style={font=\scriptsize, align=center},
  arr/.style={-{Stealth[length=2mm]}, semithick},
  dsh/.style={-{Stealth[length=2mm]}, semithick, dashed, red!70!black}
]
\node[box, minimum width=26mm] (bandit) {bandit code-selector\\(Thompson sampling)};
\node[box, anchor=west] (encA) at ($(bandit.east)+(14mm,9mm)$) {RLC\\encoder};
\node[box, anchor=west] (encB) at ($(bandit.east)+(14mm,-9mm)$) {RLC\\encoder};
\node[box, right=8mm of encB] (il) {interleaver\\($D{=}16$)};
\node[box, fill=gray!12, minimum height=14mm, anchor=west] (ch)
      at ($(il.east |- bandit)+(14mm,0)$) {channel:\\AWGN $+$ $k$ on/off\\interferers};
\node[box, right=16mm of ch] (dec) {GRAND\\decoder};
\node[box, below=15mm of dec] (nm) {noise model\\(iid\,/\,ORBGRAND\,/\,HMM)};
\node[box] (learn) at (bandit |- nm) {Bayesian\\learner};
\node[box, below=10mm of nm] (stats) {statistics\\collector};
\draw[arr] ($(bandit.east)+(0,3mm)$) -- (encA.west)
      node[pos=0.45, above, sloped, lbl] {arm A};
\draw[arr] ($(bandit.east)+(0,-3mm)$) -- (encB.west)
      node[pos=0.45, below, sloped, lbl] {arm B};
\draw[arr] (encA.east) -- ($(ch.west)+(0,4mm)$);
\draw[arr] (encB) -- (il);
\draw[arr] (il.east) -- ($(ch.west)+(0,-4mm)$);
\draw[arr] (ch) -- (dec) node[midway, above, lbl] {$y^n$};
\draw[arr] (nm) -- (dec) node[midway, right, lbl] {query order};
\draw[arr] (dec.east) -- ++(19mm,0) |- (stats.east)
      node[pos=0.3, right, lbl] {deviations $d_t$};
\draw[arr] (stats.west) -| (learn.south)
      node[pos=0.28, above, lbl] {aggregate statistics (feedback link)};
\draw[dsh] (learn.north) -- (bandit.south)
      node[midway, right, lbl] {channel estimate $\hat\theta$};
\draw[dsh] (learn.east) -- (nm.west)
      node[midway, above, lbl] {channel estimate $\hat\theta$};
\draw[dsh] (dec.north) -- ++(0,15mm) -| (bandit.north)
      node[pos=0.35, above, lbl] {ACK/NACK $\rightarrow$ reward $u$};
\coordinate (sepx) at ($(il.east)!0.5!(ch.west)$);
\draw[dashed, gray!70]
      ($(sepx |- dec.north)+(0,10mm)$) -- ($(sepx |- stats.south)+(0,-2mm)$);
\node[lbl, gray!90, anchor=north east]
      at ($(sepx |- stats.south)+(-2mm,0)$) {\emph{encoder side}};
\node[lbl, gray!90, anchor=north west]
      at ($(sepx |- stats.south)+(2mm,0)$) {\emph{decoder side}};
\end{tikzpicture}
\caption{System block diagram. Arm A is the non-interleaved RLC$(128,105)$;
arm B passes $D=16$ codewords of the same code through a cross-codeword
block interleaver. The GRAND decoder queries codebook membership in the
order supplied by the active noise model. The statistics collector
summarizes received-sample deviations $d_t$; the aggregate statistics cross
the feedback link to the learner, whose estimate $\hat\theta$ is forwarded
to the code-selector and to the receiver noise model. Decoding outcomes
provide the bandit reward.}
\label{fig:system}
\end{figure*}

Figure~\ref{fig:system} shows the closed loop. Throughout, a
\emph{codeword} is one length-$n$ RLC output, a \emph{packet} is the
transmission of one codeword and is the unit of reward accounting, and an
\emph{interleaver block} is $D$ codewords transmitted through the
interleaver. Table~\ref{tab:params} collects the implementation parameters.

\begin{table}[t]
\centering
\caption{Implementation parameters. None of the values below were tuned by
sweeping against the reported metrics; sensitivity to $c_\ell$ and to the
channel parameters is discussed in Section~\ref{sec:conclusion}.}
\label{tab:params}
\footnotesize
\setlength{\tabcolsep}{3pt}
\begin{tabular}{llp{3.1cm}}
\toprule
Parameter & Value & Selection rationale\\
\midrule
code $(n,k)$ & $(128,105)$ & high-rate regime benchmarked in the GRAND
literature; $2^{-23}$ random-hit floor\\
interleaver depth $D$ & 16 & smallest power of two $\ge \max_i T_i$\\
query budget $Q$ & $10^6$ (V0), $10^5$ (loop) & runtime; V0 validates the
budget--BLER relation\\
feedback period & 200 packets & several periods per expected convergence\\
run threshold $\tau$ & $1.5\hat\sigma$ & detection/false-run tradeoff
(Sec.~\ref{sec:estimation})\\
minimum run length & 3 & suppresses false runs from noise\\
TS discount $\gamma$ & 0.99 & effective memory $\approx 100$ packets\\
pseudo-count $m_0$ & 30 & fraction of one bandit memory\\
prediction batch & 64 packets, $Q{=}2{\times}10^4$ & per-arm Monte Carlo
cost cap\\
latency weight $c_\ell$ & 0.005 & sets delay penalty to $\approx 10\%$ of
maximum goodput\\
inform threshold & $w \ge 0.15$ & skip predictions from empty estimates\\
\bottomrule
\end{tabular}
\end{table}

\subsection{Codes and Arms}

Both arms use the same fixed realization of a systematic random linear code
with parameters $(n,k)=(128,105)$, rate $R = k/n \approx 0.82$, and
generator $G=[I_k \mid P]$ with $P$ drawn once, uniformly over
$\mathbb{F}_2^{k\times(n-k)}$. The arms differ only in whether groups of
$D=16$ codewords are interleaved before transmission. The $n-k=23$
redundancy bits put the probability that a random word passes the
membership test at $2^{-23}\approx 1.2\times 10^{-7}$ per query, negligible
at the error rates studied. Arm A (non-interleaved) transmits one codeword
per packet. Arm B (interleaved) writes $D$ codewords as rows of a
$D\times n$ array and transmits column-wise, so a contiguous interference
burst of length $T \le D$ contributes at most one error to any codeword. No
codeword in the block can be decoded until all $Dn$ symbols arrive, so
every arm-B packet carries a decoding latency of $D=16$ packet durations.

Two accounting conventions follow from the interleaver. First, when the
bandit selects arm B it commits to an entire interleaver block: the next
$D$ packets are transmitted on arm B, and the $D$ per-codeword rewards are
credited when the block is decoded. Second, the simulation applies the
interleaving delay only through the reward penalty defined in
Section~\ref{sec:bandit}; the sampler's posterior updates occur when the
block is decoded, and no additional feedback-transport delay is modeled.

\subsection{Channel}

Transmission is BPSK with unit symbol energy, $s_t = 1-2b_t$, over AWGN of
variance $\sigma^2 = 1/(2R\cdot 10^{E_b/N_0[\mathrm{dB}]/10})$. In addition,
$k$ interferers operate on a global symbol clock, asynchronously to our
packets. Interferer $i$ alternates between OFF periods of duration
$\mathrm{Exp}(\bar{L}_i)$ symbols and ON periods of exactly $T_i$ symbols;
while ON it adds $c\, a_i$ to every received sample, where the amplitude
$a_i$ is fixed for the connection and the sign $c\in\{\pm 1\}$ is drawn
once per burst and held for the burst duration. The received sample is
\begin{equation}
y_t = s_t + \textstyle\sum_{i=1}^{k} b_{i}(t)\, c_i(t)\, a_i + w_t,
\qquad w_t \sim \mathcal{N}(0,\sigma^2),
\label{eq:channel}
\end{equation}
where $b_i(t)\in\{0,1\}$ is interferer $i$'s on/off process. The
transmitter and receiver know none of $k$, $\{a_i\}$, $\{T_i\}$,
$\{\bar{L}_i\}$.

Table~\ref{tab:channel} lists the realized parameters for the reference
seed. Two features of this regime drive the study. First, bursts are much
shorter than the codeword ($T_i \le 12 \ll n=128$), so several bursts can
strike one codeword. Second, the amplitudes are comparable to the signal,
so a burst does not flip every covered bit: the flip probability of a
covered bit is $\tfrac12[Q((1-a)/\sigma)+Q((1+a)/\sigma)]$, which evaluates
to $0.09$ to $0.44$ across the three interferers. Burst errors are
therefore mostly isolated within the interference interval rather than
contiguous. This regime has an important consequence: a hard-decision
Gilbert--Elliott model provides little gain because the induced errors
carry little run structure (Section~\ref{sec:results}). The burst remains
detectable, however, in the received-sample amplitudes, since a covered
sample is displaced by $\pm a_i$ regardless of whether the bit flips.

\begin{table}[t]
\centering
\caption{Realized channel parameters (reference seed), $E_b/N_0=6$~dB,
$\sigma=0.391$, background flip probability $Q(1/\sigma)=0.0053$.}
\label{tab:channel}
\begin{tabular}{lccc}
\toprule
 & interf.~1 & interf.~2 & interf.~3\\
\midrule
amplitude $a_i$ & 0.626 & 1.148 & 1.443\\
ON length $T_i$ (sym) & 4 & 6 & 12\\
mean OFF $\bar{L}_i$ (sym) & 633 & 454 & 691\\
duty cycle & 0.6\% & 1.3\% & 1.7\%\\
flip prob.\ within burst & 0.09 & 0.32 & 0.44\\
\bottomrule
\end{tabular}
\end{table}

\subsection{Receiver}

The receiver runs GRANDAB with query budget $Q$ and one of four
interchangeable pattern generators: (i) \emph{iid}, Hamming-weight order;
(ii) \emph{Markov}, a Gilbert--Elliott run-structure order parameterized by
the collapsed interference statistics; (iii) \emph{ORBGRAND},
logistic-weight order on reliability ranks \cite{duffy2022orbgrand}; and
(iv) the \emph{learned HMM} model of Section~\ref{sec:noisemodel}. After
each decoded packet the receiver updates a statistics collector, and every
200 packets it flushes a compact feedback message on an error-free,
rate-limited feedback link. Decoding outcomes (ACK/NACK) reach the
transmitter and form the bandit reward.

In simulation, success is defined by equality of the decoded and
transmitted information words. In a deployed system this signal would
require a CRC or a higher-layer acknowledgment, since GRAND codebook
membership alone does not distinguish a correct decode from an undetected
codeword error; at the studied rates the undetected-error contribution is
bounded by the $2^{-23}$ per-query random-hit floor.

\section{Online Interference Estimation}
\label{sec:estimation}

This section describes what the receiver measures, how the interference
parameters are recovered, and how estimate confidence is quantified. We
write the estimate as the typed object
\begin{equation}
\hat\theta = \big(\hat\sigma,\ \hat k,\
\{\hat a_i,\ \hat p_i\}_{i=1}^{\hat k},\ \hat T\big),
\label{eq:thetahat}
\end{equation}
where $\hat p_i$ is interferer $i$'s estimated duty cycle and $\hat T$ a
common ON-duration estimate, and we write $w \in [0,1]$ for the associated
confidence score defined in Section~\ref{sec:confidence}. The decoder model
constructed from the estimate is denoted $\mathcal{M}(\hat\theta)$.

\subsection{Feedback Statistics}

From successfully decoded packets the receiver reconstructs the transmitted
BPSK symbols $\hat s_t$ and forms deviations $d_t = y_t - \hat s_t$, which
by \eqref{eq:channel} contain only interference plus noise. For the
interleaved arm the decoded codewords are re-interleaved first, so
statistics are always collected in channel order, where bursts are
contiguous. The feedback message contains: a signed histogram of $d_t$; a
robust noise-scale estimate $\hat\sigma$ (median absolute deviation of the
bulk); histograms of the lengths of \emph{impaired runs}, defined as
maximal runs of at least $3$ consecutive same-sign deviations exceeding a
threshold $\tau = 1.5\hat\sigma$; and, for every impaired run, a histogram
of the run's mean deviation together with the run's symbol count.

\subsection{Run-Mean Estimator}
\label{sec:runmean}

A direct estimator fits a Gaussian mixture to the per-symbol deviation
histogram: the mixture has a dominant $\mathcal{N}(0,\sigma^2)$ component
and components at $\pm a_i$ with weights equal to the duty cycles. In the
studied regime this estimator does not reliably separate the weak
interference component from the noise: with $a_1 = 0.63$ and
$\sigma = 0.39$ the component sits $1.6\sigma$ from a bulk holding $96\%$
of the mass, and converged expectation-maximization (EM) fits from multiple
initializations returned similar merged solutions, for example two
components near $(0.99, 1.50)$ for true amplitudes $(0.7, 1.3)$. This
behavior suggests that the difficulty is statistical rather than solely an
initialization failure of EM.

Run means improve the separation. A detected impaired run of length $L$
from a burst of amplitude $a$ has mean deviation concentrated near $a$ with
standard deviation $\sigma/\sqrt{L}$, at least $\sqrt{3}$ times narrower
than the per-symbol components, and the noise bulk is largely absent by
construction (only rare false runs of pure noise survive the same-sign,
length-$\ge 3$ filter). Detection introduces a bias of known form: each
symbol admitted to a run exceeds $\tau$, so it is distributed as
$\mathcal{N}(a,\sigma^2)$ truncated to $(\tau,\infty)$, with mean
\begin{equation}
g(a) \;=\; a + \sigma\,
\frac{\phi\!\big((\tau-a)/\sigma\big)}{1-\Phi\!\big((\tau-a)/\sigma\big)},
\label{eq:trunc}
\end{equation}
where $\phi$ and $\Phi$ are the standard normal density and distribution
functions. Since $g$ is strictly increasing, an observed cluster mean
$\bar m$ inverts to $\hat a = g^{-1}(\bar m)$ by one-dimensional root
finding. Equation~\eqref{eq:trunc} also locates the false-run cluster at
$g(0)$, so the mixture fit can anchor one component there.

\subsection{Parameter Recovery}

Each feedback period the learner folds the signed run-mean histogram about
zero (burst signs are symmetric) and fits a small Gaussian mixture by EM
with free means and widths, plus one component with mean fixed at $g(0)$ to
absorb false runs. Surviving component means are inverted through
\eqref{eq:trunc}; recovered atoms explained by subset sums of already-found
atoms are removed (two simultaneously active interferers produce a cluster
at $a_i + a_j$); the mixture weight of each atom, corrected by its
per-symbol detection probability, yields its duty cycle $\hat p_i$; and the
ON length $\hat T$ is the mode of the run-length histogram. On synthetic
data with well-separated amplitudes the estimator recovers all parameters
within $15\%$; amplitudes closer than about $1.5\sigma$ are recovered as a
set rather than individually, which the decoder model tolerates.

\subsection{Confidence Criterion}
\label{sec:confidence}

Let $S$ denote the accumulated run count supporting the dominant recovered
atom, and let $c$ denote the number of consecutive refits for which the
well-supported atoms moved by less than $0.15$ in Hausdorff distance. The
continuous confidence score is
\begin{equation}
w \;=\; \min\!\left(1, \tfrac{S}{100}\right)\cdot
\min\!\left(1, \tfrac{c+1}{3}\right),
\label{eq:conf}
\end{equation}
clipped to $w=0$ when no atoms exist and set to $w=1$ once the hard
criterion holds: at least three feedback periods, $c \ge 2$, and $S > 100$.
The hard criterion defines the \emph{confidence switch} used as a reference
time in the experiments.

\subsection{Why We Call the Learner Bayesian}
\label{sec:bayesjust}

We use the term \emph{Bayesian} for this learner in a specific sense, which
we state to avoid overclaiming. First, the quantity the receiver ultimately
consumes is a posterior: given $\mathcal{M}(\hat\theta)$, the
forward--backward recursion of Section~\ref{sec:noisemodel} computes exact
posterior marginals of the hidden interference state, and the resulting
$p_t$ are posterior bit-error probabilities. Second, the estimate is
represented and consumed distributionally rather than as a bare point: the
pair $(\hat\theta, w)$ functions as a posterior summary, with $w$ acting as
a precision surrogate, and both of its consumers treat it that way. On the
decoder side, the gating policies of Section~\ref{sec:gating} mix the HMM
posterior with an interference-free posterior in proportion to $w$. On the
encoder side, the bandit code-selector consumes the same posterior summary:
the transmitter computes a utility based on the learned distribution, the
predicted per-arm utility $\hat u(\text{arm} \mid \hat\theta)$ of
\eqref{eq:uhat}, by Monte Carlo against the channel law
$\mathcal{M}(\hat\theta)$, and the informed sampler of
Section~\ref{sec:informed} injects these utilities into its conjugate
Normal-Gamma posterior updates with pseudo-counts proportional to $w$, so
that the selector's belief over arm utilities inherits both the location and
the precision of the learner's estimate. Third, the point estimation itself
has a maximum a
posteriori structure: the anchored mixture component encodes exact prior
knowledge of the false-run location $g(0)$, and the exponential forgetting
applied to the accumulated statistics corresponds to a prior that
parameters may drift. What the learner does \emph{not} do is maintain
calibrated posterior distributions over $(\hat k, \{\hat a_i\})$
themselves; $w$ in \eqref{eq:conf} is a heuristic precision summary, not a
derived posterior variance. Two refinements would close this gap. A
Dirichlet-process mixture over the run means would make $\hat k$ a
posterior quantity and yield credible intervals for the amplitudes, and a
variational treatment of the anchored mixture would produce an approximate
posterior covariance from which $w$, the gating blend weight, and the
pseudo-observation counts of Section~\ref{sec:informed} could all be
derived rather than hand-designed. Both fit the existing interfaces, since
every consumer already accepts a (location, precision) pair.

\section{Learned Noise Model for GRAND}
\label{sec:noisemodel}

\subsection{HMM Construction}
\label{sec:hmm}

Given a confident estimate, the receiver builds a hidden Markov model over
the instantaneous interference level with signed states
$\{0, \pm\hat a_1, \ldots, \pm\hat a_{\hat k}\}$, transition rates derived
from $(\hat p_i, \hat T)$, and emission density
$\tfrac12\sum_{x=\pm1}\mathcal{N}(y;\,x + s,\ \hat\sigma^2)$ in level state
$s$. The signed states reflect the channel's per-burst sign convention.

For computational tractability this state space assumes a single active
interferer at a time: simultaneous activations, which produce levels such
as $\pm(a_i + a_j)$, are not represented explicitly, and their effect is
absorbed into the closest amplitude state. The approximation is accurate in
the studied regime because the learned duty cycles are $0.6$--$1.7\%$, so
the probability that two interferers overlap on a given symbol is
approximately $\sum_{i<j} p_i p_j \approx 4\times10^{-4}$, i.e., about one
symbol in $2{,}400$. The estimator's subset-sum removal
(Section~\ref{sec:estimation}) prevents these rare overlap events from
creating spurious amplitude atoms. Channels with substantially higher duty
cycles would require overlap states.

\subsection{Posterior Flip Probabilities}

For each received block, forward--backward smoothing yields state
posteriors $\gamma_t(s)$ and hence a per-bit posterior flip probability
\begin{equation}
p_t \;=\; \sum_{s} \gamma_t(s)\;
\Pr\!\big[\text{hard decision wrong} \mid y_t, s\big].
\label{eq:pt}
\end{equation}
The mechanism in \eqref{eq:pt} is that a strong deviation at symbol $t$
raises the posterior that a burst covers the neighborhood of $t$, which
raises $p_{t'}$ for neighboring bits even when $y_{t'}$ itself is close to
a nominal constellation point.

\subsection{Pattern Ordering}
\label{sec:ordering}

HMM smoothing produces correlated posterior error events, and a list of
marginals $\{p_t\}$ does not by itself define a joint pattern probability.
We therefore approximate the posterior error vector as conditionally
independent with marginals $p_t$. Under this approximation a candidate
pattern $\mathbf{z}$ has cost
\begin{equation}
\sum_{t: z_t = 1} \log\frac{1-p_t}{p_t},
\label{eq:cost}
\end{equation}
and the SGRAND max-heap \cite{solomon2020sgrand} enumerates patterns in
increasing cost, i.e., in decreasing probability under the factorized
posterior model. The ordering is exact with respect to this factorized
model, not with respect to the full joint HMM posterior.

\subsection{Receiver Gating Policies}
\label{sec:gating}

The estimate flows to the receiver every feedback period, so the remaining
design choice is when to act on it. We compare three policies.
\emph{Hard}: adopt $\mathcal{M}(\hat\theta)$ only after the confidence
switch. \emph{Eager}: adopt the raw estimate at full weight from the first
period in which any amplitude atom exists. \emph{Soft}: adopt immediately
but blend the flip posteriors, $\tilde p_t = w\, p_t^{\rm HMM} +
(1-w)\, p_t^{\rm AWGN}$, using the confidence score $w$ of
\eqref{eq:conf}; here $p_t^{\rm AWGN}$ is the interference-free posterior
$1/(1+e^{2|y_t|/\hat\sigma^2})$. At $w=0$ the blend equals the soft-ML
ordering for pure AWGN, so an empty or incorrect estimate degrades to the
generic soft decoder.

In the present design only arm A receives the learned model. Arm B
continues to use ORBGRAND because the cross-codeword interleaver
substantially weakens the local temporal dependence modeled by the HMM; we
therefore evaluate the learned ordering only on arm A. A joint model that
accounts for the interleaver permutation could also be considered, but is
outside the scope of this study.

\section{Code Selection as a Bandit}
\label{sec:bandit}

The transmitter treats the two transmission modes as arms of a bandit and
receives, for each packet, the reward
\begin{equation}
u \;=\; R\cdot\mathbf{1}\{\text{decoded correctly}\}
\;-\; c_{\ell}\,\ell(\text{arm}),
\label{eq:reward}
\end{equation}
where $R$ is the code rate, $\ell$ is the decoding latency in packet
durations ($\ell=0$ for arm A, $\ell=D=16$ for arm B), and
$c_{\ell}=0.005$. The reward is a dimensionless engineering utility: $R$ is
dimensionless, $\ell$ is measured in packets, and $c_\ell$ is a design
weight that places delay on the same scale as successful delivery rather
than a conversion derived from a queueing or application model. With these
constants the arm-B penalty is $c_\ell D = 0.08$, about $10\%$ of the
maximum per-packet goodput $R=0.82$. Conclusions that depend on the
relative ranking of the arms depend on $c_\ell$; mapping the policy
boundary over $c_\ell$ is identified as future work.

Because the environment is nonstationary by construction (the learner's
retuning of arm A's decoder changes arm A's success probability
mid-connection), we use discounted Thompson sampling with a Normal-Gamma
posterior per arm. Every update first decays all posteriors toward the
prior with factor $\gamma = 0.99$, giving an effective memory of roughly
$100$ packets, and then applies the standard conjugate update to the played
arm; selection samples a mean from each posterior and plays the argmax.
This is the standard construction for switching bandits
\cite{garivier2011switching,russo2018tutorial}: the decay keeps posterior
variance from collapsing, so the sampler continues to explore at a low rate
and adapts when an arm's reward distribution changes.

\subsection{Model-Informed Thompson Sampling}
\label{sec:informed}

The ACK-only sampler learns each arm's utility from reward observations
alone, so after the decoder retuning it must re-estimate arm A's improved
success probability from new rewards. The transmitter, however, holds the
information needed to predict the change: given $\hat\theta$ it can
simulate each arm against the estimated channel. In the model-informed
variant, at each feedback period the transmitter synthesizes a channel with
parameters $\hat\theta$, runs a Monte Carlo batch (64 packets per arm,
query budget $2\times10^4$) of each arm's current decoder against it, and
forms the predicted utility
\begin{equation}
\hat u(\text{arm} \mid \hat\theta) \;=\;
R\,\hat P(\text{success} \mid \text{arm}, \hat\theta)
\;-\; c_\ell\, \ell(\text{arm}),
\label{eq:uhat}
\end{equation}
mirroring \eqref{eq:reward}. The prediction enters the sampler as
pseudo-observations: arm $i$'s Normal-Gamma posterior receives a batch
update equivalent to $m = m_0 \cdot w$ samples of value $\hat u(i)$, with
$m_0 = 30$. Real ACK/NACK evidence continues to arrive through the ordinary
path, and the discount removes stale predictions unless they are refreshed,
so the posterior is a confidence-weighted combination of model prediction
and observed reward. The anticipated benefit is a faster preference change
at the confidence switch; the anticipated risk is that \eqref{eq:uhat}
predicts mean utility from the estimated channel and would not represent
reward burstiness or an estimate that is confidently wrong.
Section~\ref{sec:informedresults} evaluates both effects.

\section{Experimental Design and Hypothesis}
\label{sec:experiments}

We validate and then compose the components in six phases, each varying one
element while holding the rest fixed (Table~\ref{tab:phases}).

\begin{table}[t]
\centering
\caption{Experimental phases. Fixed throughout: RLC$(128,105)$, BPSK,
feedback period 200 packets, $c_\ell = 0.005$, $D=16$.}
\label{tab:phases}
\footnotesize
\setlength{\tabcolsep}{3pt}
\begin{tabular}{llll}
\toprule
Phase & Channel & Varied & Fixed\\
\midrule
V0 & AWGN & $E_b/N_0 \in [3,8]$ dB & iid GRAND, $Q{=}10^6$\\
V1 & interf., 6 dB & decoder $\times$ arm & oracle $\hat\theta$, $Q{=}10^5$\\
V2 & interf., 6 dB & learner (estimator/oracle) & arm A fixed\\
V2b & interf., 6 dB & gating (hard/eager/soft) & arm A fixed\\
V3 & interf., 6 dB & seed $\in\{0..4\}$ & ACK-only TS\\
V3b & interf., 6 dB & bandit (ACK/informed) & seed $\in\{0..2\}$\\
\bottomrule
\end{tabular}
\end{table}

Phase V0 checks the decoder core against theory on AWGN. Phase V1 measures
the performance attainable when the channel parameters are known, per
decoder and arm. Phase V2 closes the loop with the estimator replacing the
oracle, arm A fixed, and measures convergence time and the gap to the
oracle baseline. Phase V2b varies only the gating policy of
Section~\ref{sec:gating}. Phase V3 activates the bandit for $20{,}000$
packets per seed, and phase V3b compares the ACK-only and model-informed
samplers on common channel realizations.

The hypothesis for V3 is as follows. With an untuned (iid or ORBGRAND)
decoder, a burst that places four or more scattered flips in one arm-A
codeword frequently exhausts the query budget, whereas interleaving spreads
the same flips over $16$ codewords, each of which decodes at low cost; the
interleaved arm should therefore earn higher reward initially. After the
confidence switch, arm A's HMM-ordered decoder should decode most burst-hit
packets within budget, so arm A should achieve comparable or lower error
rate with no latency penalty, and the sampler's preference should reverse.
The timing of the reversal should be governed by the confidence criterion
plus the sampler's discounted memory.

Unless stated otherwise, BLER values are point estimates over the stated
packet counts; where error counts are small we report the counts and a
$95\%$ Wilson interval. All rolling curves use windows of consecutive
packets as defined in each caption, and adjacent window values are highly
correlated.

\section{Results}
\label{sec:results}

\subsection{Decoder Validation on AWGN (V0)}

\begin{figure}[t]
\centering
\includegraphics[width=0.9\linewidth]{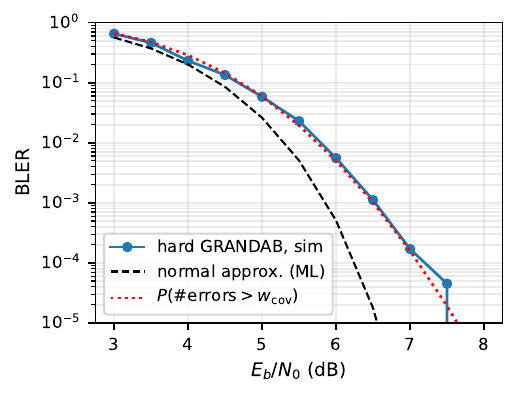}
\caption{V0: hard-detection GRANDAB on AWGN, RLC$(128,105)$, $Q=10^6$.
Each simulated point uses at least 100 block errors or $2\times10^5$
packets. Dashed: finite-blocklength normal approximation; dotted:
abandonment bound $P(\#\text{errors} > 3)$.}
\label{fig:v0}
\end{figure}

Figure~\ref{fig:v0} overlays the simulated block error rate of
hard-detection GRANDAB on AWGN against two references: the
finite-blocklength normal approximation for the induced binary symmetric
channel, and the abandonment bound $P(\#\text{errors} > w_{\rm cov})$,
where $w_{\rm cov}=3$ is the deepest error weight fully covered by the
$10^6$-query budget. The simulation follows the abandonment bound across
four decades (at $6.5$~dB, $1.11\times10^{-3}$ simulated,
$100$ errors observed, versus $1.01\times10^{-3}$ bound). In this regime
GRANDAB's error rate is set by the query budget rather than by ML
ambiguity, so the budget is the parameter through which a designer trades
worst-case work for BLER.

\subsection{Oracle Decoder Comparison (V1)}

Table~\ref{tab:v1} reports the decoder$\times$arm matrix on the
interference channel with oracle channel parameters. Three differences are
relevant. First, before any tuning, the interleaved arm has lower BLER:
B-ORBGRAND at $1.15\times10^{-1}$ versus A-ORBGRAND at
$1.51\times10^{-1}$. Second, the hard-decision Markov model improves on iid
by only about $10\%$ ($2.39\times10^{-1}$ versus $2.65\times10^{-1}$),
consistent with Section~\ref{sec:model}: the burst flips are mostly
isolated and carry little run structure. Third, the HMM model reduces arm-A
BLER from $1.51\times10^{-1}$ to $6.5\times10^{-3}$ (13 errors in 2{,}000
packets; $95\%$ Wilson interval $[3.8, 11.1]\times10^{-3}$), an absolute
reduction of $0.145$, with mean query count reduced from $17{,}296$ to
$1{,}134$. The interference is weakly expressed in the hard decisions but
detectable in the soft amplitudes, and \eqref{eq:pt} converts that
information into query order.

\begin{table}[t]
\centering
\caption{V1: BLER and mean queries per packet with oracle channel
parameters (2{,}000 packets per row, $E_b/N_0=6$ dB, $Q=10^5$). Starred
models use the oracle parameters.}
\label{tab:v1}
\small
\setlength{\tabcolsep}{4pt}
\begin{tabular}{llrrr}
\toprule
Arm & Noise model & errors & BLER & queries/pkt\\
\midrule
A (non-interl.) & iid & 530 & $2.65\times10^{-1}$ & 27{,}842\\
A (non-interl.) & ORBGRAND & 301 & $1.51\times10^{-1}$ & 17{,}296\\
A (non-interl.) & Markov$^{*}$ & 478 & $2.39\times10^{-1}$ & 25{,}360\\
A (non-interl.) & HMM$^{*}$ & \textbf{13} & $\mathbf{6.5\times10^{-3}}$ & \textbf{1{,}134}\\
B (interleaved) & iid & 477 & $2.39\times10^{-1}$ & 27{,}187\\
B (interleaved) & ORBGRAND & 230 & $1.15\times10^{-1}$ & 14{,}215\\
\bottomrule
\end{tabular}
\end{table}

\subsection{Closed-Loop Estimation (V2)}

\begin{figure}[t]
\centering
\includegraphics[width=\linewidth]{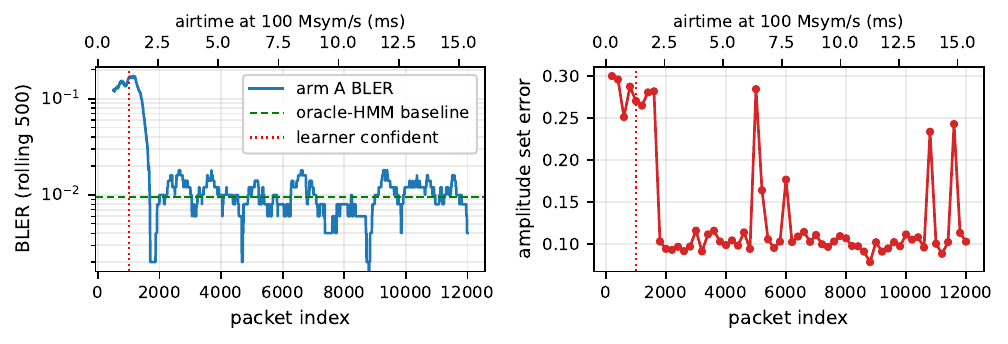}
\caption{V2: closed-loop estimation with arm A fixed (12{,}000 packets,
reference seed). Left: BLER over a rolling window of the most recent 500
packets; the dashed line is the oracle-HMM baseline measured on the same
channel realization. Right: Hausdorff distance between the estimated and
true amplitude sets at each 200-packet feedback update. The vertical line
marks the hard confidence switch; top axes give the idealized airtime
mapping of Section~\ref{sec:timescales}.}
\label{fig:v2}
\end{figure}

With the estimator replacing the oracle and arm A fixed,
Fig.~\ref{fig:v2} shows the trajectory. The estimator reaches the hard
confidence criterion after five feedback periods ($1{,}000$ packets). Arm
A's BLER is $1.43\times10^{-1}$ before the switch (143 errors / 1{,}000
packets) and $1.23\times10^{-2}$ after (135 / 11{,}000), against an
oracle-HMM baseline of $9.4\times10^{-3}$ (113 / 12{,}000) on the same
channel realization; the two post-switch intervals overlap within a factor
of $1.3$. The final estimate recovers the interferer count exactly
($\hat k = 3$) and the amplitudes as $(0.728, 1.122, 1.465)$ against truth
$(0.626, 1.148, 1.443)$. The upward bias on the weakest atom is the
run-detection bias discussed in Section~\ref{sec:runmean} at $1.6\sigma$
separation. The learned $\hat T = 4$ collapses the three ON lengths
$(4,6,12)$ to their mode; the posterior in \eqref{eq:pt} is driven mainly
by the amplitude evidence, so this coarse timing estimate suffices here.

The oracle value here ($9.4\times10^{-3}$) differs from the oracle-HMM row
of Table~\ref{tab:v1} ($6.5\times10^{-3}$) because the two runs use the
same channel parameters but independent noise and burst realizations over
different packet counts ($12{,}000$ versus $2{,}000$); the Wilson intervals
$[7.8,11.3]\times10^{-3}$ and $[3.8,11.1]\times10^{-3}$ overlap.

\subsection{Receiver Gating Policies (V2b)}
\label{sec:softgate}

\begin{figure}[t]
\centering
\includegraphics[width=0.92\linewidth]{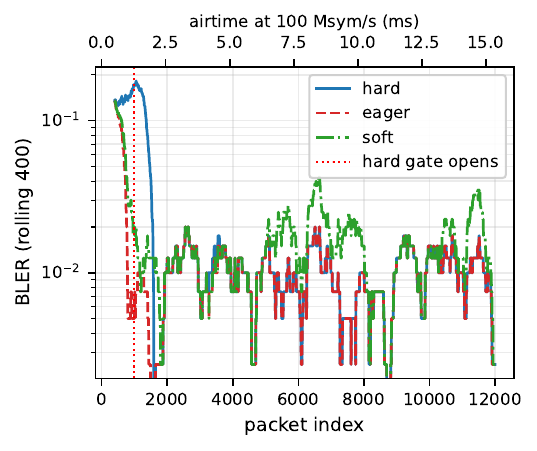}
\caption{V2b: BLER under the three gating policies for the reference seed,
computed over a rolling window of the most recent 400 packets, on a common
channel realization. The vertical line marks the first feedback update at
which the hard confidence criterion holds; the top axis gives the idealized
airtime mapping of Section~\ref{sec:timescales}.}
\label{fig:softgate}
\end{figure}

Table~\ref{tab:softgate} and Fig.~\ref{fig:softgate} compare the three
gating policies of Section~\ref{sec:gating} on common channel realizations,
with arm A fixed. Before the hard confidence threshold, eager adoption
produced the lowest BLER in all three seeds; relative to hard gating the
reduction ranged from $1.4\times$ to $4.5\times$ (absolute reductions of
$0.06$ to $0.157$). After the threshold, eager adoption was also slightly
better than hard and soft adoption in the tested runs. These results
indicate that the early estimates already contain useful ranking
information for query ordering. The soft blend is a robustness mechanism
against poor estimates; that protection did not improve performance in the
evaluated channel realizations, in which even the first-period estimates
pointed the query order at the impaired bits. The benefit of early adoption
was largest for the seed with the slowest confidence (seed 2, a
$2{,}000$-packet window). A reasonable design based on these runs is to let
the confidence criterion govern external reporting and prediction weighting
while the decoder consumes the estimate as soon as it exists, since the
blend's AWGN limit already bounds the downside of an empty estimate.

\begin{table}[t]
\centering
\caption{V2b: BLER inside the pre-confidence window (packets before the
hard gate opens, per seed) and after it, for the three gating policies. Arm
A fixed, 12{,}000 packets per run, common channel realization per seed.}
\label{tab:softgate}
\small
\setlength{\tabcolsep}{4pt}
\begin{tabular}{clccc}
\toprule
Seed & window & hard & eager & soft\\
\midrule
0 & first 1{,}000 & $1.43\times10^{-1}$ & $\mathbf{5.7\times10^{-2}}$ & $6.7\times10^{-2}$\\
  & after         & $1.23\times10^{-2}$ & $\mathbf{9.5\times10^{-3}}$ & $1.41\times10^{-2}$\\
\addlinespace[2pt]
1 & first 600     & $2.23\times10^{-1}$ & $\mathbf{1.63\times10^{-1}}$ & $1.88\times10^{-1}$\\
  & after         & $1.61\times10^{-2}$ & $\mathbf{1.29\times10^{-2}}$ & $1.97\times10^{-2}$\\
\addlinespace[2pt]
2 & first 2{,}000 & $2.01\times10^{-1}$ & $\mathbf{4.5\times10^{-2}}$ & $6.2\times10^{-2}$\\
  & after         & $1.54\times10^{-2}$ & $\mathbf{1.11\times10^{-2}}$ & $1.33\times10^{-2}$\\
\bottomrule
\end{tabular}
\end{table}

\subsection{Arm-Preference Reversal (V3)}

\begin{figure}[t]
\centering
\includegraphics[width=\linewidth]{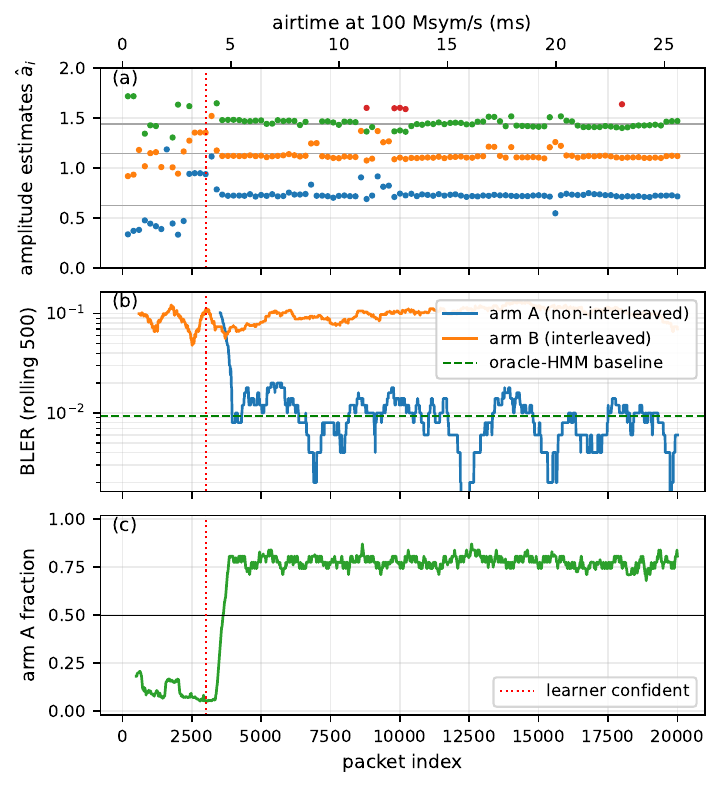}
\caption{V3 results for seed 0 over 20{,}000 packet decisions.
(a)~Estimated amplitude atoms at each 200-packet feedback update;
horizontal lines show the realized interferer amplitudes. (b)~Per-arm BLER
computed over that arm's most recent 500 transmissions, plotted at the
global packet index. (c)~Fraction of arm-A selections in a rolling
500-decision window. The vertical line marks the first update at which the
hard confidence criterion holds. The top axis is the idealized airtime
mapping of Section~\ref{sec:timescales}.}
\label{fig:timeline}
\end{figure}

Figure~\ref{fig:timeline} shows the full system for the reference seed:
per-period amplitude estimates, per-arm rolling BLER, and the sampler's
arm-A selection fraction, on a common time axis. Before the confidence
switch the sampler concentrates on arm B (arm-A fraction $0.11$); the
sampler's preference for arm B also slows statistics collection relative to
V2, since failed packets contribute no statistics, which is why confidence
arrives at packet $3{,}001$ here versus $1{,}000$ in V2. At the switch, arm
A's BLER decreases by an order of magnitude within about one sampler
memory, and the selection fraction crosses $0.5$ and settles near $0.78$;
the residual arm-B share is the discounted sampler's exploration. Over the
whole run arm A produced BLER $1.26\times10^{-2}$ (168 errors / 13{,}328
packets) at mean reward $0.810$, and arm B produced $9.41\times10^{-2}$
(628 / 6{,}672) at mean reward $0.663$.

Table~\ref{tab:seeds} reports all five seeds. The same preference reversal
occurs in every run: the post-confidence arm-A fraction lies between $0.76$
and $0.82$, compared with $0.11$ to $0.29$ before confidence, and arm A's
BLER is $5\times$ to $10\times$ lower than arm B's in each run.

\begin{table}[t]
\centering
\caption{V3 across seeds (20{,}000 packets each): packet index of the hard
confidence switch, arm-A selection fraction before and after the switch,
and per-arm BLER over the full run.}
\label{tab:seeds}
\small
\setlength{\tabcolsep}{4pt}
\begin{tabular}{crcccc}
\toprule
Seed & Conf.\ (pkt) & \multicolumn{2}{c}{arm A fraction} &
\multicolumn{2}{c}{BLER}\\
 & & before & after & arm A & arm B\\
\midrule
0 & 3{,}001 & 0.11 & 0.78 & $1.3\times10^{-2}$ & $9.4\times10^{-2}$\\
1 & 3{,}800 & 0.17 & 0.82 & $2.4\times10^{-2}$ & $1.9\times10^{-1}$\\
2 & 1{,}809 & 0.22 & 0.82 & $1.9\times10^{-2}$ & $2.0\times10^{-1}$\\
3 & 1{,}401 & 0.29 & 0.76 & $7.5\times10^{-3}$ & $6.3\times10^{-2}$\\
4 & 4{,}207 & 0.29 & 0.80 & $2.4\times10^{-2}$ & $1.6\times10^{-1}$\\
\bottomrule
\end{tabular}
\end{table}

\subsection{ACK-Only versus Model-Informed Selection (V3b)}
\label{sec:informedresults}

\begin{figure}[t]
\centering
\includegraphics[width=0.92\linewidth]{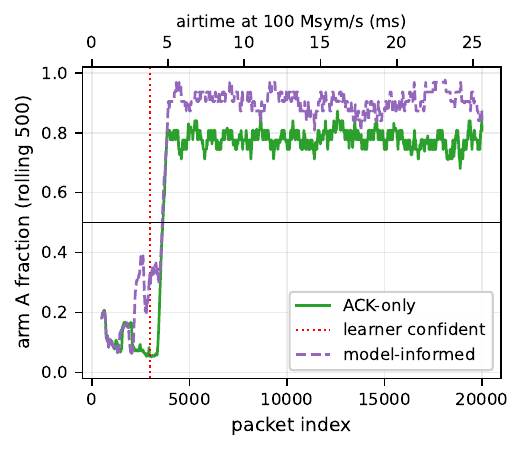}
\caption{V3b: arm-A selection fraction in a rolling 500-decision window
under ACK-only and model-informed Thompson sampling, reference seed, common
channel realization. The vertical line marks the ACK-only run's hard
confidence switch; the top axis gives the idealized airtime mapping of
Section~\ref{sec:timescales}.}
\label{fig:informed}
\end{figure}

Table~\ref{tab:informed} and Fig.~\ref{fig:informed} compare the ACK-only
sampler with the model-informed variant on common channel realizations. The
model-informed sampler reduces both the transition delay and the
post-transition selection rate of arm B. The ACK-only sampler completes the
preference reversal within $254$--$412$ packets of the confidence switch
(roughly one discounted memory), and the informed variant within
$163$--$256$ packets, a $23$--$38\%$ reduction. The larger effect is
sustained concentration: refreshed pseudo-observations keep each arm's
posterior mean close to its predicted utility with reduced variance, so the
post-switch share of arm-B selections decreases from $0.18$--$0.22$ to
$0.06$--$0.10$. The estimate-derived predictions also raise arm-A play
before confidence ($0.20$--$0.31$ versus $0.11$--$0.22$). Over the full
runs, mean reward increases by $+0.014$ to $+0.026$ per packet
($2$--$3.5\%$). In these runs the Monte Carlo predictions did not steer the
sampler to the lower-reward arm at any feedback period; we note that
\eqref{eq:uhat} predicts mean utility only, and that a confidently
incorrect estimate would inject misleading pseudo-observations. The
confidence weighting and the continuing ACK/NACK stream are the safeguards
evaluated here; drift scenarios that stress them are left for future work.

\begin{table}[t]
\centering
\caption{V3b: ACK-only versus model-informed Thompson sampling (20{,}000
packets, common channel per seed). Transition delay is measured from the
hard confidence switch to a sustained arm-A majority in a rolling
100-decision window.}
\label{tab:informed}
\footnotesize
\setlength{\tabcolsep}{3pt}
\begin{tabular}{clccc}
\toprule
Seed & Bandit & trans.\ delay (pkt) & arm-B share post & mean reward\\
\midrule
0 & ACK-only & 412 & 0.221 & 0.7610\\
  & informed & \textbf{256} & \textbf{0.095} & \textbf{0.7746}\\
\addlinespace[2pt]
1 & ACK-only & 287 & 0.183 & 0.7345\\
  & informed & \textbf{220} & \textbf{0.075} & \textbf{0.7604}\\
\addlinespace[2pt]
2 & ACK-only & 254 & 0.181 & 0.7501\\
  & informed & \textbf{163} & \textbf{0.062} & \textbf{0.7705}\\
\bottomrule
\end{tabular}
\end{table}

\subsection{Error Rate and Latency Together}

The preference reversal reflects both reliability and delay. Before
adaptation, arm B has lower BLER at the cost of a $16$-packet decoding
delay, and the sampler's reward comparison favors it. After adaptation, arm
A has both lower BLER (by $5\times$ to $10\times$ across seeds) and zero
interleaving delay. Interleaving is beneficial here only while the decoder
treats the errors as approximately independent; after the receiver exploits
temporal correlation directly, the interleaved arm retains its delay cost
but no longer provides a reliability advantage. This is the sequential,
learned counterpart of the fixed-channel comparison in \cite{an2022bursts}.
In a fuller system model the arm-B block delay would also postpone ACK
feedback and lengthen any retransmission loop; as noted in
Section~\ref{sec:model}, the present simulation represents the delay only
through the reward penalty.

\section{Equivalent Airtime at Wideband Symbol Rates}
\label{sec:timescales}

The packet axis converts to time once a symbol rate is fixed. We emphasize
that the following is an idealized airtime conversion, not an end-to-end
latency estimate: it assumes continuous allocation of the full symbol
stream to the simulated link and excludes control overhead, reference
signals, scheduling delay, feedback transport, and sharing with other
users.

A 5G~NR carrier at $100$~MHz with $30$~kHz subcarrier spacing carries $273$
resource blocks, i.e.\ $3{,}276$ subcarriers by $28$ OFDM symbols per
millisecond, roughly $9.2\times10^{4}$ resource elements per millisecond,
equivalent to about $92$~Msym/s of modulated symbols. Treating the BPSK
stream as occupying $100$~Msym/s, one $128$-symbol packet occupies
$1.28$~$\mu$s. Under this mapping: one feedback period ($200$ packets) is
$256$~$\mu$s; estimator convergence ($1.4$k--$4.2$k packets) corresponds to
$1.8$--$5.4$~ms of occupied symbol time; the preference reversal completes
within roughly $1$--$3$~ms after that; and the interleaver's
$Dn = 2{,}048$-symbol block depth corresponds to $26$~$\mu$s of added
decoding delay.

Two caveats bound this conversion. First, if packets are scheduled
sparsely, for example one per $0.5$~ms slot, the same packet counts
correspond to $0.7$--$2.1$~s; the estimator's currency is packets observed,
not seconds elapsed. Second, the mapping assumes the interference process
is stationary over the transient; interferers whose amplitudes or duty
cycles drift on comparable timescales would require the discounted
estimator to track rather than converge, which its forgetting factor is
designed for but which we have not evaluated. At high symbol rates the
estimator accumulates packet observations quickly, so the learning
transient studied here is short in absolute terms precisely in wideband
deployments.

The millisecond-scale transient also suggests where this mechanism could
sit relative to existing 5G adaptation loops, all of which operate on
comparable timescales. Link adaptation via CQI-indexed MCS tables adjusts
rate to average SINR; the code-structure choice studied here is a
complementary axis, and one integration path would treat interleaved and
non-interleaved variants of an MCS entry as distinct table rows selected by
the same bandit machinery. HARQ interacts in two ways: its ACK/NACK stream
is exactly the reward signal the sampler already consumes, so no new
signaling is needed for the ACK-only variant, and the interleaved arm's
block delay stretches the HARQ round trip, so the latency term in
\eqref{eq:reward} has a concrete counterpart in retransmission timing
budgets. CSI reporting is the existing container closest to our feedback
message: interference measurement resources (CSI-IM) already carry
interference statistics to the network on few-millisecond periodicities,
and the burst-run histograms of Section~\ref{sec:estimation}, at a few
hundred bytes per 200-packet period, are of comparable weight to a rich CSI
report. We have not evaluated these integrations; the point of the
conversion above is that the learning loop is fast enough that none of
these existing control loops would need to be slowed down to accommodate
it.

\section{Conclusion and Future Work}
\label{sec:conclusion}

We studied online selection between interleaved and non-interleaved
transmission when receiver feedback is used to learn a bursty interference
process and to update a GRAND noise model during a connection. In the
evaluated configurations, the interleaved arm was preferred before receiver
adaptation, whereas the non-interleaved arm was preferred after the learned
HMM model was activated. The change occurred because the adapted decoder
could exploit temporal interference structure directly, while the
interleaved arm retained its block-delay penalty. Using partial estimates
before the hard confidence threshold reduced pre-convergence BLER in all
tested runs. Confidence-weighted model predictions also accelerated the
change in arm preference and reduced post-transition selection of the
lower-reward arm. These results support treating interleaving policy,
channel learning, and decoder adaptation as coupled design decisions.

The results are limited to one high-rate random linear code, BPSK, one
interleaver depth, one closed-loop operating SNR, and five realizations of
a synthetic on/off interference model. The control plane is idealized:
receiver-to-transmitter statistics are carried over an error-free link
without an airtime charge, and transmitter-to-receiver parameter updates
incur a fixed one-period delay but no transport cost. Statistics are formed
only from successfully decoded packets, which may bias the channel
estimate. The decoder HMM omits simultaneous interferer states, and its
GRAND ordering uses a factorized approximation to the joint posterior.
Interleaving latency is included in the reward, but the full effects of
delayed acknowledgments and retransmissions are not modeled. The present
results should therefore be interpreted as evidence for the proposed
closed-loop mechanism rather than as a general criterion for disabling
interleaving.

Future work should first characterize the operating region in which each
transmission mode is preferred. This requires sweeps over the latency
weight, SNR, code rate, block length, interleaver depth, and interference
parameters, together with drifting-channel experiments that test whether
the estimator and model-informed sampler can track stale or changing
estimates. A fuller protocol study should assign explicit bit budgets, loss
models, scheduling delays, and airtime costs to both directions of the
control link, and should include delayed ACK and HARQ interactions.

A second direction is to generalize the learned receiver. The factorized
pattern model of Section~\ref{sec:ordering} could be replaced by an
ordering that uses more of the joint HMM posterior, and soft-output
decoders such as SOGRAND \cite{yuan2025sogrand} could provide the selector
with a graded correctness signal rather than binary ACK/NACK feedback. The
heuristic confidence score could also be replaced by a posterior precision
obtained from a variational or nonparametric mixture model
(Section~\ref{sec:bayesjust}). More generally, the estimator could learn a
distribution or latent representation of the channel
\cite{miyamoto2025universal} rather than parameters of the present on/off
interference model.

Finally, the two-arm selector can be extended to structured families of
codes and transmission modes that vary rate, block length, and interleaver
depth. Such a setting would permit comparisons among Thompson sampling, UCB
\cite{auer2002ucb}, DSEE \cite{vakili2013dsee}, EXP3 \cite{auer2002exp3},
and structured or combinatorial bandit methods
\cite{gai2012combinatorial}. The current pseudo-observation update could
also be replaced by a contextual policy that conditions directly on
features of the learned channel model.

\section*{AI Use Acknowledgement}

The author has used Claude Code for writing code and simulations, and for
help with design of the learning algorithm. Several AI tools including
Claude Code, ChatGPT, Grok and Gemini were used to help with the paper
writing, editing, proofreading, and formatting. The human author accepts
full responsibility for the contents of this work.

\bibliographystyle{IEEEtran}
\bibliography{refs}

\end{document}